\documentclass[12pt]{article}
\usepackage[utf8]{inputenc}
\usepackage[T1]{fontenc}
\usepackage[english]{babel}
\usepackage{lmodern, csquotes, eurosym}
\usepackage[colorlinks=true, citecolor=blue, linkcolor=blue, urlcolor=blue, pdfborder={0 0 0}]{hyperref}
\usepackage[round]{natbib}
\usepackage{etoolbox}

\bibpunct{(}{)}{;}{a}{}{,}

\makeatletter

\patchcmd{\NAT@citex}
  {\@citea\NAT@hyper@{%
     \NAT@nmfmt{\NAT@nm}%
     \hyper@natlinkbreak{\NAT@aysep\NAT@spacechar}{\@citeb\@extra@b@citeb}%
     \NAT@date}}
  {\@citea\NAT@nmfmt{\NAT@nm}%
   \NAT@aysep\NAT@spacechar\NAT@hyper@{\NAT@date}}{}{}

\patchcmd{\NAT@citex}
  {\@citea\NAT@hyper@{%
     \NAT@nmfmt{\NAT@nm}%
     \hyper@natlinkbreak{\NAT@spacechar\NAT@@open\if*#1*\else#1\NAT@spacechar\fi}%
       {\@citeb\@extra@b@citeb}%
     \NAT@date}}
  {\@citea\NAT@nmfmt{\NAT@nm}%
   \NAT@spacechar\NAT@@open\if*#1*\else#1\NAT@spacechar\fi\NAT@hyper@{\NAT@date}}
  {}{}

\makeatother

\makeatother

\makeatletter
\def\@fnsymbol#1{
  {\ifcase#1\or
    *\or
    \S\or
    \ddagger\or
    \P\or
    \|\or
    **\or
    \dagger\dagger\or
    \ddagger\ddagger\else
    \@ctrerr
  \fi}}
\makeatother

\usepackage{tikz}
\usepackage{caption}
\usepackage{subcaption}
\usepackage{mathtools}
\usepackage{comment}
\usepackage{booktabs}  
\usepackage{siunitx}

\usepackage{amssymb}
\usepackage{amsfonts}
\usepackage{geometry}
\usepackage{graphicx}

\title{AI Recommendations and Non-instrumental Image Concerns\thanks{I am especially grateful to Yuval Salant, Jörg Spenkuch, Alex Imas, Daniel Martin, Alvaro Sandroni, Benjamin Golub, and Tali Dayan for their continuous support. I also thank Effi Benmelech, Joshua Dean, Uri Gneezy, Anders Humlum, Rafael Jiménez-Durán, Annie Liang, Ryan Oprea, Devin Pope, and Avner Strulov-Shlain, as well as the audiences at the Booth Behavioral Economics Lab and the Kellogg Strategy Brown Bag Seminar, for their helpful comments. The experiment was pre-registered on aspredicted.org (\#220425). The IRB ID at Northwestern University for the experiment is STU00222514.}}
\author{David Almog\thanks{Kellogg School of Management, Northwestern University, \href{mailto:david.almog@kellogg.northwestern.edu}{david.almog@kellogg.northwestern.edu}.}}
\date{\today}
\medskip

\begin{document}

\renewcommand{\thefootnote}{\fnsymbol{footnote}}
\maketitle
\renewcommand{\thefootnote}{\arabic{footnote}}

\begin{abstract}
There is growing enthusiasm about the potential for humans and AI to collaborate by leveraging their respective strengths. Yet in practice, this promise often falls short. This paper uses an online experiment to identify non-instrumental image concerns as a key reason individuals underutilize AI recommendations. I show that concerns about how one is perceived, even when those perceptions carry no monetary consequences, lead participants to disregard AI advice and reduce task performance.
\end{abstract}


\newpage 

\section{Introduction}
We are witnessing a surge in the use of artificial intelligence (AI) systems across various work environments, a trend likely to intensify in the coming years. As algorithmic predictive capabilities advance and digitization and data collection efforts expand, new opportunities are emerging to observe how professionals make decisions with the assistance of these tools. Machine learning algorithms are eclipsing experts in many areas, including bail judges predicting pretrial misconduct \citep{Kleinberg2018}, radiologists predicting pneumonia from chest X-rays \citep{Rajpurkar2017, Topol2019}, and workforce professionals predicting productivity for hiring and promotion \citep{Chalfin2016}. While AI has the potential to outperform many human professionals, there is hope that human-AI collaboration can yield even better results by leveraging the unique strengths of each. AI excels at processing large volumes of data and remains free from emotional biases, while humans may possess private information or be better equipped to handle edge cases. 

One of the most popular decision-making frameworks involves humans making choices based on AI recommendations, a structure that preserves decision-making authority in human hands. However, making decisions in environments with algorithmic recommendations requires humans to exercise discretion—an area that research shows can be particularly challenging. For instance, \citet{Hoffman2018} highlights the difficulties in exercising discretion within hiring contexts, where managers often overrule algorithmic recommendations based on personal biases rather than legitimate private information. Similarly, \citet{Agarwal2023} finds that providing radiologists with access to AI predictions does not, on average, improve performance, primarily due to factors such as overconfidence and correlation neglect. In a recent study of a pretrial bail court, \citet{Angelova2023} report that 90\% of judges perform worse than an algorithm when they override its recommendations. The implication is striking: nearly every judge in that setting could improve decision quality by always following the algorithmic advice. So why don’t they? The answer may lie not in a lack of information or incentives, but in deeper psychological forces that shape how people engage with algorithmic tools.

In this paper I explore a new mechanism that contributes to the explanation on why human-AI collaboration, particularly in AI recommendation systems, often falls short of expectations. This paper focuses on studying the impact of non-instrumental image concerns (also referred to as \textit{hedonic motives} by \citet{Bursztyn2017}) on AI utilization rates. Are workers worried about how others perceive them when they use AI? If so, image concerns may discourage workers from making decisions based solely on accuracy. For instance, a worker who consistently follows AI recommendations may be perceived as exerting low effort or contributing little value \citep{Weitzner2024}, while one who rarely follows them may appear resistant to technological adaptation. These examples illustrate that image concerns could plausibly push AI utilization in either direction. However, recent evidence suggests that social norms are beginning to emerge against the use of AI. For example, \citet{Ling2025} find that social desirability bias leads people to under-report their AI usage. Similarly, \citet{Yanagizawa-Drott2025} show that human evaluators tend to penalize LLM-generated job applications, even though this is not formally specified as an evaluation criterion. While image concerns may affect payoffs either directly or indirectly, my focus here is to show that they can also reflect psychological forces that operate independently of monetary incentives.

I conducted an online experiment in Prolific where participants completed a 50-round image classification task with assistance from AI recommendations. Incentives were tied solely to performance: one round was randomly selected for payment, and participants earned a bonus if their final answer for that round was correct. At the end of the study, all participants were required to join a brief video call with a member of the research team to learn whether they had earned the bonus. For those randomly assigned to the treatment group, the video call also included a short review of the participant’s AI usage, comparing it to the average behavior of others. This intervention was designed to trigger image concerns without introducing reputational or monetary consequences.

While such concerns may be even stronger in real-world settings—where people care more deeply about how they are perceived by peers, supervisors, or close collaborators—cleanly isolating non-instrumental image concerns in the field is methodologically challenging. The controlled online experimental setting used here allowed me to shut down instrumental channels and induce a relatively modest image concern shock. Nonetheless, the effect was meaningful: participants in the treatment group relied significantly less on AI and performed worse, despite being incentivized solely on accuracy. Specifically, AI usage dropped by 4.5 percentage points (10\%), and accuracy declined by 2.7 percentage points (3.3\%). These findings underscore how subtle psychological forces, such as concern about how one's choices are perceived by a recently met individual, can distort decision making, even at the cost of task success.

One of the most important questions in the AI adoption discourse is who stands to benefit—and who may be left behind. With this in mind, the paper examines whether the image concerns studied here affect specific segments of the population differently. The heterogeneity analysis reveals that minority participants are significantly more affected by image-related pressures, exhibiting stronger treatment effects in the form of lower AI adoption and reduced performance relative to white participants. 

Before concluding the study, participants were asked to report their level of discomfort regarding the upcoming video call. The results support the presence of psychological forces: participants in the treatment group were 55 percent more likely to report feeling uncomfortable about the call. They also expressed lower satisfaction with how effectively they used the AI recommendations during the task, which is consistent with the idea that their choices were shaped by additional constraints related to image concerns.

There are two main reasons to believe the findings from this online experiment could generalize horizontally across a range of field settings. First, the study captures participants’ responses to being observed by someone from the research team they had just met. In most real-world work environments, however, image concerns are shaped by interactions with colleagues, clients, or other stakeholders with whom individuals have ongoing relationships and more meaningful reputational stakes, which we could expect to amplify the effects observed here. Second, the image categorization task is not the type of activity typically associated with skill or intelligence, whereas many workplace tasks are more explicitly tied to cognitive ability. This stronger association could increase image-related pressures in other contexts.

The remainder of the paper is structured as follows. Section \ref{sec:lit} reviews the relevant literature. Section \ref{sec:Experiment} describes the experimental design and discusses key design choices. Section \ref{sec:Results} presents the main results on AI utilization and task performance, followed by an analysis of heterogeneous effects and complementary evidence on the underlying mechanism. Section \ref{sec:Con} concludes.

\subsection{Related Literature}\label{sec:lit}

First and foremost, this paper contributes to the growing literature on human-AI collaboration. While AI systems continue to improve their predictive capabilities, there remain many domains in which humans stay in the decision loop. This persistence can be attributed to several factors: collaborative synergies that enhance efficiency, social preferences rooted in tradition or ethical concerns, and labor market frictions that slow full automation. One format where human and AI collaboration has delivered promising results is with Large Language Models \citep{Brynjolfsson2023, Noy2023, Otis2024, Peng2023}, which is commonly best suited for writing or articulation tasks. Another collaboration structure involves humans making the initial decision, and AI has the ability to overrule them - referred to as \textit{AI oversight} by \citet{Almog2025}. They exploit a unique context with tennis umpires to show that AI oversight can improve performance, as long as we account for the underlying psychological forces involved in being corrected in public.\footnote{In a companion paper, \citet{Almog2025b} examine how workers respond differently to AI oversight compared to human oversight.} Yet another collaboration structure commonly used in practice involves humans making decisions based on AI recommendations, which has the desirable feature of preserving final decision rights in human hands. This study examines the role of image concerns using the decision-making framework described last. 

High-stakes professional environments have faced the perils and challenges of implementing AI recommendations. Medical clinics \citep{Agarwal2023}, judicial courts \citep{Angelova2023}, and HR offices \citep{Hoffman2018} are just a few examples of settings where the benefits of accessible AI recommendations have yet to materialize. A common denominator across these studies is the systematic under-utilization of AI recommendations, a phenomenon previously coined as \textit{algorithmic aversion} \citep{Dietvorst2015}. The most common explanation for these underwhelming outcomes centers on overconfidence and other deviations from Bayesian updating. Using controlled experiment environments, \citet{Caplin2024} provide evidence that calibration and belief formation play a key role in determining the gains from working with AI recommendations, while \citet{Serra-Garcia2023} show that the timing of algorithmic feedback matters: receiving feedback before forming one's own beliefs significantly increases adoption. If belief formation explanations alone were sufficient to account for the issue, it would be reasonable to assume that professionals only care about maximizing accuracy. However, there is already evidence suggesting that working with AI may influence the preferences of the parties involved.\footnote{Specifically in the context of recommendation systems, \citet{McLaughlin2024} propose a model of reference-dependent utility in which the reference point is influenced by the AI recommendation. \citet{Albright2024} provides empirical evidence suggesting that algorithmic recommendations may offer judges reputational cover when making visible mistakes. In an AI oversight setting, \citet{Almog2025} show that being overruled by Hawk-Eye in public asymmetrically altered the costs of different error types for professional tennis umpires.} \citet{Mills2022} provide one of the few bright spots in the economics literature, documenting collaboration gains when humans operate under AI recommendations. Their study examines a setting in which Child Protective Services workers had to decide whether to send an investigator to screen a reported family. A key implementation feature was that workers were encouraged by their managers to consult the AI tool during team meetings and received training to understand their comparative advantage relative to the algorithm.\footnote{Workers were aware that the algorithm was trained only on family background information, while details from the call report were treated as “private information” available only to the worker.} These distinctions in the implementation may have helped reduce the stigma associated with using the tool, an important factor that may not have been addressed in other settings where human-AI collaboration failed to generate complementarities. I present evidence that non-instrumental image concerns may help reconcile previously mixed empirical findings and offer guidance on the environments under which human–AI collaboration yields better outcomes.

Image concerns can influence consequential personal decisions, including voting \citep{Dellavigna2017}, credit consumption \citep{Bursztyn2018}, educational investment \citep{Bursztyn2019}, political contributions \citep{Perez-Truglia2017}, and childhood immunization \citep{Karing2024}. This study adds to the growing body of research on image concerns (for a review, see \citet{Bursztyn2017}) by identifying a new domain in which these forces shape behavior: human–AI collaboration. Image concerns are also deeply relevant in the workplace, where non-monetary incentives play a central role. Workers often derive utility not just from wages, but also from meaning \citep{Cassar2018, Ashraf2025}, pride, respect \citep{Ellingsen2007, Ellingsen2008}, and how they are compared to fellow workers \citep{Mas2009}. This paper’s findings that non-instrumental image concerns can deter AI utilization support the view that non-monetary aspects of the workplace can meaningfully affect workers’ decisions.

This paper also contributes to the understanding of stigmatized behaviors \citep{Celhay2025, Friedrichsen2018}, which are naturally intertwined with social image concerns. A closely related paper is \citet{Golub2019}, whose main contribution is to unpack the mechanisms behind the stigma of seeking information by disentangling its signaling and shaming components. They find that signaling is the primary driver of stigma. In my setting, participants behave as if there is a similar stigma associated with relying on AI recommendations. While this study is framed within the context of human-AI collaboration, another key distinction from \citet{Golub2019} is that I deliberately shut down the instrumental reputational channel to focus on the non-instrumental dimension of image concerns.

Although decision-making under recommendations is a longstanding form of collaboration, the use of AI as the recommending agent introduces additional considerations. In a study of digital payments in the Senegalese taxi industry, \citet{Houeix2025} shows that the close link between data observability and digital technologies can slow down adoption. In the same spirit, AI recommendations create by design a digital trace that is both durable and easily auditable. This makes AI recommendations particularly susceptible to image concerns—unlike advice from, say, a co-worker in a meeting, where attribution is more diffuse and harder to corroborate. Thus, the present work also adds to the literature on the economics of digitalization by identifying a channel through which digital systems can facilitate the emergence of image-related pressures.

\section{Experiment Design} \label{sec:Experiment} 
\subsection{Task}
The study consisted of completing 50 rounds of an image classification task. Each round, participants observed a blurred image and had to choose the correct category from a list of 16 options.\footnote{Airplane, bear, bicycle, bird, boat, bottle, car, cat, chair, clock, dog, elephant, keyboard, knife, oven, and truck.} After making their choice, participants were shown an AI recommendation and given the option to switch their initial answer to match the AI recommendation, if they preferred to. The task can thus be conceived as a two-stage decision process: in the initial stage, participants select the category they believe is correct; in the second stage, after receiving the AI recommendation, they have the option to revise their initial answer and switch to the recommended one. The dynamics of this two-stage decision process are illustrated in Appendix \ref{TaskInterface}, while Appendix \ref{sec:TaskAppendix} provides detailed information on how the task was constructed.

Image categorization falls within the broader class of data annotation tasks in which human input remains fundamental, as is the case with most supervised learning processes. Even in some unsupervised learning cases, such as ChatGPT, humans contribute during the later stages of reinforcement learning. So, image categorization is particularly useful because it allows for experimental control, while closely resembling a common gig task performed by a segment of the online workforce.

\subsection{Implementation}
In a pre-registered online experiment, I recruited 220 U.S. residents via Prolific to complete an image classification task.\footnote{Preregistration plan available at https://aspredicted.org/8sq9-hw8w.pdf.} Participants received a \$7 base payment, with the opportunity to earn up to \$12. A \$5 bonus award was determined solely by performance and chance: one round was randomly selected, and participants earned the bonus if their final answer (after reviewing the AI recommendation) was correct. There were no monetary incentives specifically for using the AI, except that using it well could lead to more correct answers and a higher chance of earning the bonus. The average payment was \$11.2, well above Prolific’s minimum of \$4 for studies of comparable duration (30 minutes).

Participants were randomly assigned to one of two experimental conditions (110 participants in each condition). For all participants, the video call served the purpose of revealing which round had been selected for payment and whether they had earned the bonus. The only difference was that participants in the treatment group were informed that the research team member conducting the call would also review their AI usage during the task and compare it to the average participant’s behavior.\footnote{As suggested by \citet{Haaland2023}, providing statistical information about others' behavior can help participants contextualize and interpret their own use of AI. What’s important here is that the treatment has already taken effect by the time the video call occurs, so the key is not the actual comparison but the moment participants anticipate it. Once the video call dynamic is explained, participants can easily project themselves into a situation where their AI use might be compared to others'.} AI usage was explicitly defined as the percentage of rounds in which participants changed their answer to match the AI recommendation. All participants were told that the video call would be brief, lasting no more than 1 to 2 minutes, and that they were not expected to engage in any way beyond listening to their results. Appendix \ref{Instructions} shows screenshots of the instructions and how they differ by treatment, where applicable.

After completing the 50 rounds of the task, participants answered two non-incentivized questions about their attitude toward the video call and their perception of how effectively they utilized the AI recommendations during the task. The results of these questions are reviewed in the discussion of potential mechanisms.

Finally, participants were given a link to join the video call hosted in a private Zoom meeting room. They were instructed to change their Zoom username to their Prolific ID to maintain anonymity and ensure accurate matching of participation outcomes. Participants were first directed to an online waiting room and admitted one at a time to manage the flow and ensure that no two participants joined simultaneously. As outlined in the instructions, I personally hosted all video calls and adhered strictly to the script corresponding to each participant’s assigned experimental condition.

Following best practices outlined by \citet{Haaland2023}, the intervention is brief and delivered in neutral language, ensuring that the explanation of the AI review carries neither positive nor negative connotations. It is further supported by a graphical illustration designed to help participants project themselves into the moment the call occurs.

Although \citet{deQuidt2018} find only modest evidence of experimenter demand effects (EDE) in typical online experiments, I carefully consider the potential impact of EDE when discussing the experimental design in Section \ref{sec:ExperimentDesign}, outlining which types of EDEs pose a threat to the experiment and which are less concerning.

\subsection{Video Call}
One key element of the recruitment teaser was the explicit requirement that participants complete a video call with video and audio enabled at the conclusion of the study. Participants were asked to join the study only if they were willing to complete the call.

At the beginning of the study, participants were reminded that participation in the video call was required and were given another opportunity to return their submission if they were unable or unwilling to join. They were also informed that the video call would be used to reveal whether they earned the \$5 bonus, which could not be claimed without completing the call.

I aimed to be as straightforward as possible in the recruitment materials to avoid attrition from participants unwilling to join a video call as part of the study. This likely resulted in a non-representative sample of the broader Prolific participant pool. If anything, the study may have excluded those most sensitive to the treatment, suggesting that the observed effects could be even stronger in the general population. In practice, attrition was relatively low: only 3.6\% of participants (5 in the treatment group and 3 in the control group) completed the task but either were unable or chose not to join the final video call.

Outside the study context, image concerns are likely stronger because people interact with individuals they know and expect to encounter again. While the absence of that dynamic in an experimental setting helps rule out instrumental image concerns arising from expectations of future interactions, it also makes non-instrumental concerns harder to induce. Implementing the intervention through a video call provides a modest but meaningful boost to a force such as non-instrumental image concerns, which can be challenging to manipulate experimentally.

\subsection{AI Recommendations}
In each round, participants received an AI recommendation after making their initial choice. When the AI recommendation differed from their original answer, they had the opportunity to revise their response and switch to the one suggested by the AI. Participants were informed that the recommendations came from an AI model trained on the same task and that it was accurate 85\% of the time, a statement that holds true for both the full dataset and the subset of images used in the experiment.

Participants were also told that the AI might perform well on images that humans find difficult but could occasionally miss images that seem easy to the human eye. This statement reflects a genuine feature of the original study of \citet{Steyvers2022}. Prior research \citep{Dietvorst2015,Dreyfuss2025} shows that people may overreact to a single poor algorithmic recommendation, so this disclaimer was intended to prevent participants from losing trust in the AI based on an isolated error.

\subsection{Design Choices}\label{sec:ExperimentDesign}

In this section, I discuss several of the main choices faced while designing the experiment.

\textit{Between-subject.} Participants are exposed to only one experimental condition to avoid priming effects. A between-subjects design helps mitigate the most pressing form of experimenter demand, which \citet{Zizzo2010} describes as a task construal cue—a purely cognitive type of EDE. While a within-subjects design could yield valuable insights into the distribution of treatment effects (particularly in a setting like this, where effects may vary in direction across individuals), it also poses a significant risk of triggering an EDE as a direct response to the experimental manipulation.

\textit{Categorization task.} This task and particular dataset offer several advantages. First, data labeling/annotation requires human involvement, as it is essential for AI training and currently cannot be fully automated. Second, it is a well-established and popular category on online labor platforms. Third, it is easy to explain and requires minimal to no training, especially for platform users with previous labeling experience. Fourth, it provides excellent experimental control: the dataset includes ground-truth labels, contains variations in image difficulty (measured by noise level), and has been designed for testing human-AI collaboration. Finally, the images already come with AI predictions, which makes it convenient to use real AI output in the experiment and ensures consistency with the instructions provided to the participants.

\textit{Recommendation timing.} Providing the recommendation after an initial choice allows us to clearly identify the AI's impact on the final decision. This approach avoids ambiguity in cases where the initial choice and the recommendation align, as it would otherwise be impossible to determine whether the decision was influenced by the AI or made independently. This design has two key advantages: first, it enables precise measurement of how often recommendations are followed; second, it conveys to participants that their choices to follow recommendations are being reliably tracked.\footnote{In some work environments, identifying which employees rely more heavily on recommendations, when these are provided before making an initial choice, can take considerable time (e.g., days or even months). However, this study does not account for extended observation over such type of timeframes.}

\textit{Incentives.} By paying a bonus based on the outcome of a randomly selected round, we clearly convey that monetary rewards are only related to performance. This incentive scheme allows us to attribute any treatment effects to a non-instrumental utility component.

\textit{Video call treatment.} A video call provides the most credible intervention to genuinely convey that another person is observing participants' AI usage; other alternatives, such as a personalized message, are less persuasive given how easily they can be automated today. Face-to-face interactions provide the strongest opportunity to experimentally induce image concerns in a one-time encounter with a stranger. This method has been widely used in experimental research to isolate the non-instrumental dimension of image concerns, demonstrating that interacting with a stranger can influence behaviors such as voting \citep{Dellavigna2017} and charitable giving \citep{Dellavigna2012}. Another option was to host the experiment in person at the university lab. However, it is difficult to ignore the student–experimenter connection in that setting, as students who participate in lab studies often do so repeatedly and may develop an ongoing relationship with the lab.

\textit{Experimenter hosting the call.} The primary reason for having a member of the research team host the video call is to create an interaction between the participant and someone who appears to genuinely care and is well-positioned to form inferences about their behavior. This setup helps replicate a key feature of real-world environments, where workers assisted by AI recommendations are influenced by the judgments of others involved. To the extent that participants form beliefs about the experimenter—who plays the role of an authority figure—and respond accordingly, this could be considered a \textit{social pressure} type of EDE (following the terminology of \citet{Zizzo2010}). However, this type of EDE is not a concern as a confound, as long as the interaction is conducted in neutral language, because such responses are precisely the object of study.

\section{Results}\label{sec:Results}

Table \ref{tab:Descriptive} provides a summary of participants’ characteristics as reported by Prolific, including gender, age, ethnicity, and prior experience on the platform (measured by the number of approved studies). Two-sided t-tests indicate that treatment assignment was balanced across the observable characteristics available.

\begin{table}[h!]
    \centering
    \caption{Descriptive Statistics and Balance Check.}
    \label{tab:Descriptive}
    \sisetup{table-format=1.3}  
    \begin{tabular}{lSSSSSS}
        \toprule
        & {Control} & & {Treatment} & \phantom{space} & \multicolumn{2}{c}{Comparison} \\
        \cmidrule(lr){2-2} \cmidrule(lr){4-4} \cmidrule(lr){6-7}
        & {Mean (sd.)} &  & {Mean (sd.)} & & {Difference} & {t-stat}   \\
        \midrule
        Male & {0.50 (0.048)}  & & {0.45 (0.048)} & & {0.05} & {0.67} \\
        Age & {41.85 (1.33)}  & & {42.45 (1.43)} & & {-0.6} & {-0.31} \\
        Minority & {0.4 (0.047)}  & & {0.36 (0.046)} & & {0.04} & {0.55} \\
        100+ Approvals & {0.84 (0.035)}  & & {0.86 (0.033)} & & {-0.02} & {-0.56} \\
        1000+ Approvals & {0.38 (0.047)}  & & {0.41 (0.047)} & & {-0.03} & {-0.41} \\
        
        \bottomrule
    \end{tabular}
    
    \vspace{0.5em}
    \begin{minipage}{0.9\textwidth} 
        \footnotesize \textit{Notes}: 110 participants per experimental condition. Minority is a dummy variable that takes the value of one for participants identifying with any non-white ethnicity.
        \\
        * \(p<0.10\), ** \(p<0.05\), *** \(p<0.01\).
    \end{minipage}
\end{table}

\subsection{Treatment Effects}

Even though monetary incentives were solely tied to task performance and did not differ across experimental conditions, participants in the treatment group, who were told their AI usage would be reviewed by a member of the research team, responded in line with the two pre-registered hypotheses: lower AI utilization and lower performance.

The empirical analysis relies on the following simple regression framework: 

\begin{equation}\label{eqn:EmpiricalSpecificationMain}
Y_{ij}= \alpha + \beta{T_i} + \gamma X_j + \epsilon_{ij}
\end{equation}

$Y_{ij}$ denotes the outcome variable for participant $i$ and image $j$. $T_i \in \{0,1\}$ is an indicator equal to 1 if participant $i$ was assigned to the treatment group. $X_j$ represents image fixed effects. The coefficient of interest is $\beta$, which measures the average treatment effect of being assigned to the treatment group. All the results include image fixed effects and standard errors clustered at participant level.\footnote{Results are robust without the image fixed effects.} Treatment effects for the six pre-registered outcome variables are reported in Table \ref{tab:Results}. 

Columns 1 and 2 report the treatment effects on the use of AI recommendations. Column 1 includes all rounds, while Column 2 includes only rounds in which participants had the opportunity to switch from their initial answer to the AI recommendation, meaning the recommendation differed from their initial choice. It is important to note that when the AI recommendation matched the participant’s initial choice, this does not count as following the AI. Across both metrics, we observe a statistically significant decrease in the use of AI recommendations. When considering all rounds, participants in the treatment group followed the AI recommendation 4.5 percentage points (p.p.) less often, representing a 10\% decrease from the 44\% rate observed in the control group. If we focus only on the rounds where participants actually have the option to follow the AI recommendation and forgo their original answer, the estimated effect is 6.9 p.p., a 9\% decrease from the 79\% rate from the control group.

\begin{table}[h!]
    \centering
    \caption{Treatment Effects.}
    \label{tab:Results}
    \sisetup{table-format=1.3}  
    \begin{tabular}{lSSSSSSSSS}
        \toprule
        & \multicolumn{2}{c}{AI Recommendation use} & & \multicolumn{2}{c}{Correct answer} & & \multicolumn{2}{c}{Response time} \\
        \cmidrule(lr){2-3} \cmidrule(lr){5-6} \cmidrule(lr){8-9}
        & {All} & {Conditional} & & {Initial} & {Final} & & {Initial} & {Rec. stage}  \\
        & {(1)} & {(2)} & & {(3)} & {(4)} & & {(5)} & {(6)}   \\
        \midrule
        \textbf{Treatment} & \textbf{-0.045}** & \textbf{-0.069}*** & &\textbf{0.005} & \textbf{-0.027}**  &  & \textbf{1.38} & \textbf{-0.568} \\\vspace{1em}
        & {(0.021)} & {(0.026)} & & {(0.018)} & {(0.011)}  & & {(1.24)}  & {(0.576)}   \\

        Constant & 0.440 & 0.787 & & 0.468 & 0.809  & & 14.54 & 6.68 \\
        Observations & {11,000} &  {6,098} &  & {11,000} & {11,000} & &{11,000} &  {6,098} \\
        \bottomrule
    \end{tabular}
    
    \vspace{0.5em}
    \begin{minipage}{0.9\textwidth} 
        \footnotesize \textit{Notes}: Each dependent variable is regressed on a treatment group indicator (equal to 1 if the participant was assigned to the treatment group), with image-specific fixed effects included. Standard errors are clustered at the participant level.\\
        * \(p<0.10\), ** \(p<0.05\), *** \(p<0.01\).
    \end{minipage}
\end{table}

\begin{figure}[htbp]
\centering
\caption{CDF on Average Recommendation Use per Participant.}
\begin{subfigure}[b]{.49\textwidth}
\centering
\includegraphics[width=3.2in]{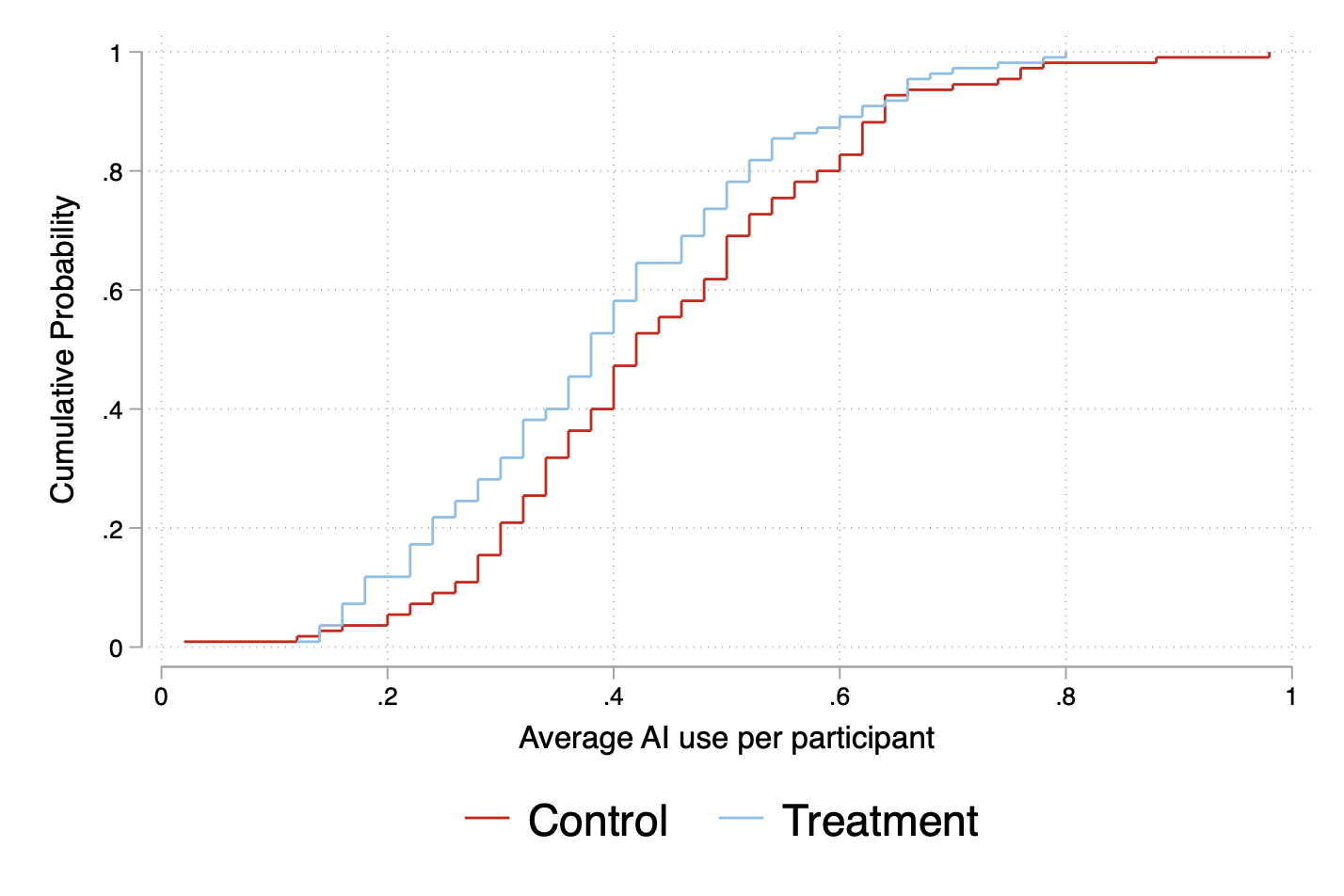}
\caption{All rounds.}
\end{subfigure}
\begin{subfigure}[b]{.49\textwidth}
\centering
\includegraphics[width=3.2in]{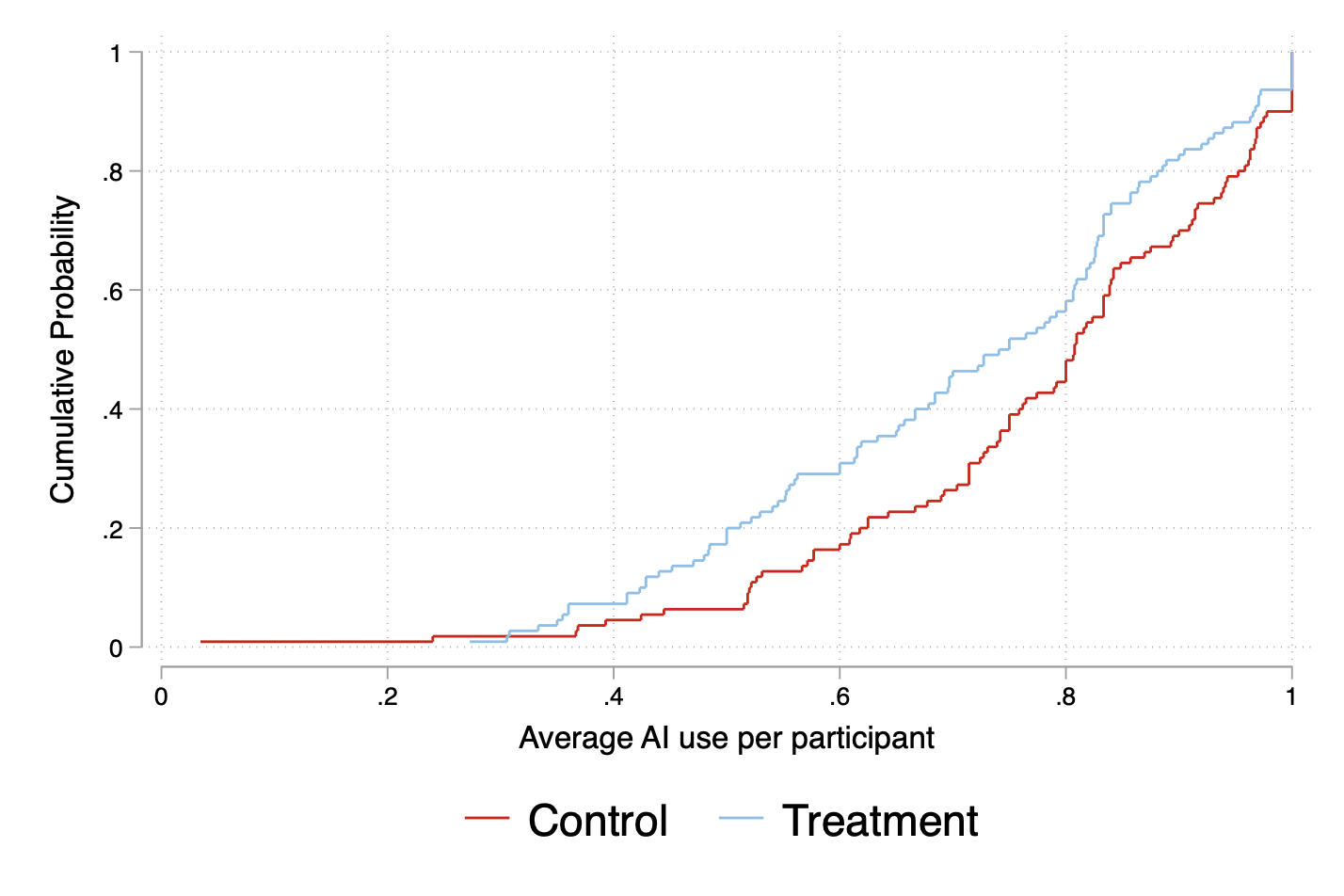}
\caption{Conditional on switching available.}
\end{subfigure}

    \vspace{0.5em}
    \begin{minipage}{0.9\textwidth} 
        \footnotesize \textit{Notes}: The one-sided Kolmogorov–Smirnov test evaluating whether AI use values in the control group are larger than those in the treatment group yields the following p-values: 0.129 for the left figure and 0.012 for the right figure.
    \end{minipage}
\label{fig:CDFs}
\end{figure}

To better understand how the treatment reduced AI use, Figure \ref{fig:CDFs} plots the empirical cumulative distribution of average AI use per participant, separately by treatment group. Regardless of the definition used for following an AI recommendation (whether based on all rounds or only those where switching was possible), the distribution for the control group first-order stochastically dominates that of the treatment group. Because this is a between-subjects analysis, it is not possible to determine whether the distributional shift is driven by large changes among a subset of participants or by a small, systematic decrease across the board. However, the result clearly reflects a first-order stochastic dominance (FOSD) shift. The one-sided Kolmogorov–Smirnov test lacks power when using all rounds (p-value = 0.129), but yields significant results when restricting to rounds in which switching answers was possible (p-value = 0.012).

The treatment effects on AI utilization do not emerge solely from the initial rounds (immediately after the instructions emphasized the upcoming call) or the final rounds (just before the call occurred). As shown in Figure \ref{fig:10rounds} in the Appendix, the effects are consistent throughout the 50 rounds. This pattern suggests that the observed treatment effects are not purely driven by the two most salient moments of the intervention.

The main reason to care about non-instrumental image concerns reducing AI utilization is that decision-makers are leaving accurate recommendations on the table when they choose to rely on their own judgment instead. This ultimately leads to lower performance by failing to fully leverage the tools available. In this experiment, the reduction in AI use among participants in the treatment group translates into a measurable decline in performance. Performance was measured by whether the participant’s final answer—after reviewing the AI recommendation and completing the round—was correct. This was the outcome participants were financially incentivized for, and it is also the metric that objectively matters in real-world applications. Going back to Table \ref{tab:Results}, Column 4 present the treatment effects on the final correct answer, the average correct response rate was 80.9\% in the control group, compared to 78.2\% in the treatment group. This reflects a 2.7 p.p. (3.3\%) decrease in performance, significant at the 5\% level. A simple back-of-the-envelope calculation that rationalizes the observed 2.7 p.p. drop in performance—alongside the reduction in following AI recommendations of 4.5 p.p.—suggests that treated participants substituted AI recommendations, which were known to be 85\% accurate, with their own answers that were correct only about 25\% of the time.\footnote{This result comes from solving for the participant’s own accuracy, $x$, that would generate a 2.7 p.p. drop in performance when participants rely on their own answers instead of AI recommendations 4.5 percent of the time. This relationship is captured in the following equation: $4.5*(0.85)-2.7=4.5*(x)$.} These results indicate that participants are rejecting AI recommendations in favor of inferior personal judgments. By deliberately overriding the AI, they make their task objectively harder. This behavior aligns with findings from the task selection literature, where workers sometimes choose more difficult tasks to signal expertise \citep{Siemsen2008,Katok2011} or to manage self-image in the presence of ego \citep{Köszegi2006}. Column 6 in Table \ref{tab:Results} shows no significant treatment effect on response time during rounds in which participants had the option to revise their answer. This supports the interpretation that there were no shifts in effort during the recommendation revision stage.

Since the option to revise an answer was only offered when the AI recommendation differed from the participant’s initial choice, a valid concern is that the treatment may have influenced behavior at the initial choice stage. Specifically, the treatment could have prompted participants to exert more effort initially, leading them to encounter a systematically different set of images for which the AI recommendation would later be shown.\footnote{\cite{Agarwal2025} Illustrate how effort responses should be considered when designing human-AI collaboration systems.} Having this in mind, I deliberately selected a task with minimal learning potential and limited returns to additional effort, aiming to ensure that the revision stage of the decision process remained as comparable as possible across treatment conditions. Regressing initial correct answers on Response time (controlling for image-level noise and participant fixed effects), the estimated effect is a 2.2 p.p. decrease in accuracy per additional 10 seconds spent. The effect is not only small considering the average response time was 14 seconds, but also negative. This indicates that higher performance in this task tends to come from faster decisions, reinforcing the idea that effort offers limited returns. Finally, Columns 3 and 5 of Table \ref{tab:Results} provide evidence against a distortion in initial choice behavior, showing no significant treatment effects on either accuracy or response time at the initial choice stage.

\subsection{Heterogeneity Analysis}
Among the most pressing challenges for policymakers and institutions integrating AI into decision-making processes is the risk that these technologies may reinforce existing societal biases, disproportionately affecting vulnerable populations and exacerbating social inequities \citep{Gillis2021}. It is difficult to pinpoint which specific life experiences or cultural factors might underlie non-instrumental image concerns in a given setting like the one studied in this paper. However, it is plausible that these factors are correlated within social groups, leading to systematic differences in how individuals respond to being observed while using AI.

As an exploratory analysis, I examine heterogeneous effects across gender, age, and ethnicity using the following empirical specification:

\begin{equation}\label{eqn:EmpiricalSpecification}
Y_{ij}= \alpha + \beta_1{T_i} + \beta_2{G_i} + \beta_3T_i\times G_i + \gamma X_j + \epsilon_{ij}
\end{equation}

Where $Y_{ij}$ denotes the outcome variable for participant $i$ and image $j$. $T_i\in\{0,1\}$ is an indicator variable equal to 1 if participant $i$ was assigned to the treatment group. $G_i\in\{0,1\}$ is an indicator equal to 1 if participant $i$ belongs to the group of interest (e.g., male, older, or minority, depending on the analysis). The interaction term $T_i\times G_i$ captures the differential treatment effect for the group of interest relative to the omitted group, thus the associated coefficient $\beta_3$ is the primary parameter of interest. $X_j$ represents image fixed effects. All standard errors are again clustered at the participant level.

Table \ref{tab:Results_Heterogeneity} in the Appendix reports the estimated coefficients from the heterogeneity analysis described above. There is no significant difference in AI utilization or accuracy by gender or age (where “older” is defined as age 40 or above, approximately the median in the sample). Using a large-scale survey in Denmark, \citet{Humlum2024} document a gender gap in ChatGPT adoption. However, the null result found here by gender does not support the hypothesis that such a gap is driven by non-instrumental image concerns. In contrast, despite reduced statistical power, the findings for minority participants are striking: when pooling all non-white participants, the treatment effect reveals an additional 9.6 p.p. (23\%) reduction in AI recommendation use relative to white participants, significant at the 5\% level.\footnote{When conditioning on rounds where participants had the option to follow the AI recommendation, the additional effect for minorities is a reduction of 11.1 p.p. (14.4\%).} This decline in AI utilization among minorities is mirrored in their performance: their final accuracy rate falls by 4 p.p. (5\%) relative to white participants, significant at the 10\% level.

\begin{figure}[htbp]
\centering
\caption{Heterogeneous Treatment Effects by Ethnicity.}
\begin{subfigure}[b]{.49\textwidth}
\centering
\includegraphics[width=3.1in]{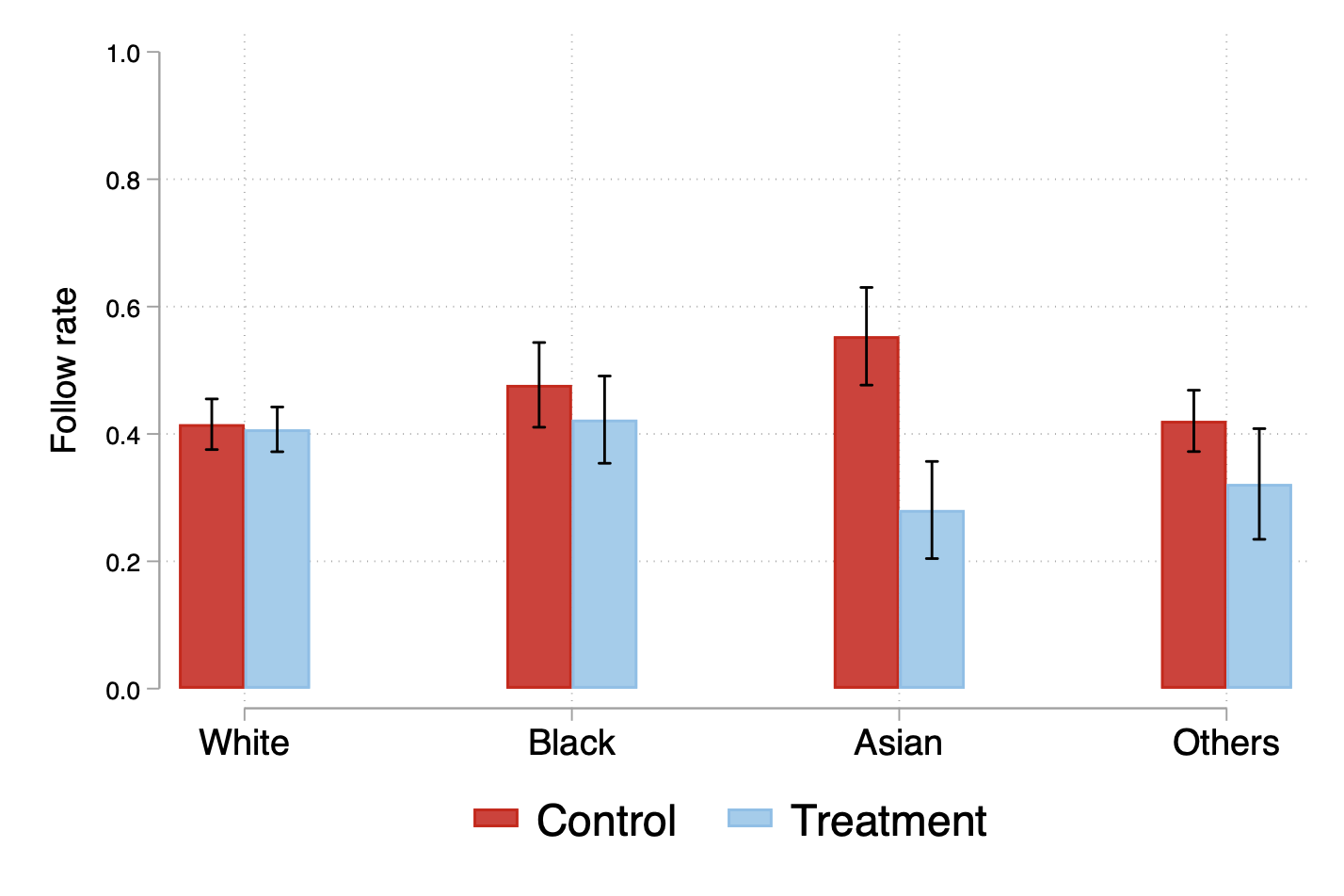}
\caption{AI recommendations use (all rounds).}
\end{subfigure}
\begin{subfigure}[b]{.49\textwidth}
\centering
\includegraphics[width=3.1in]{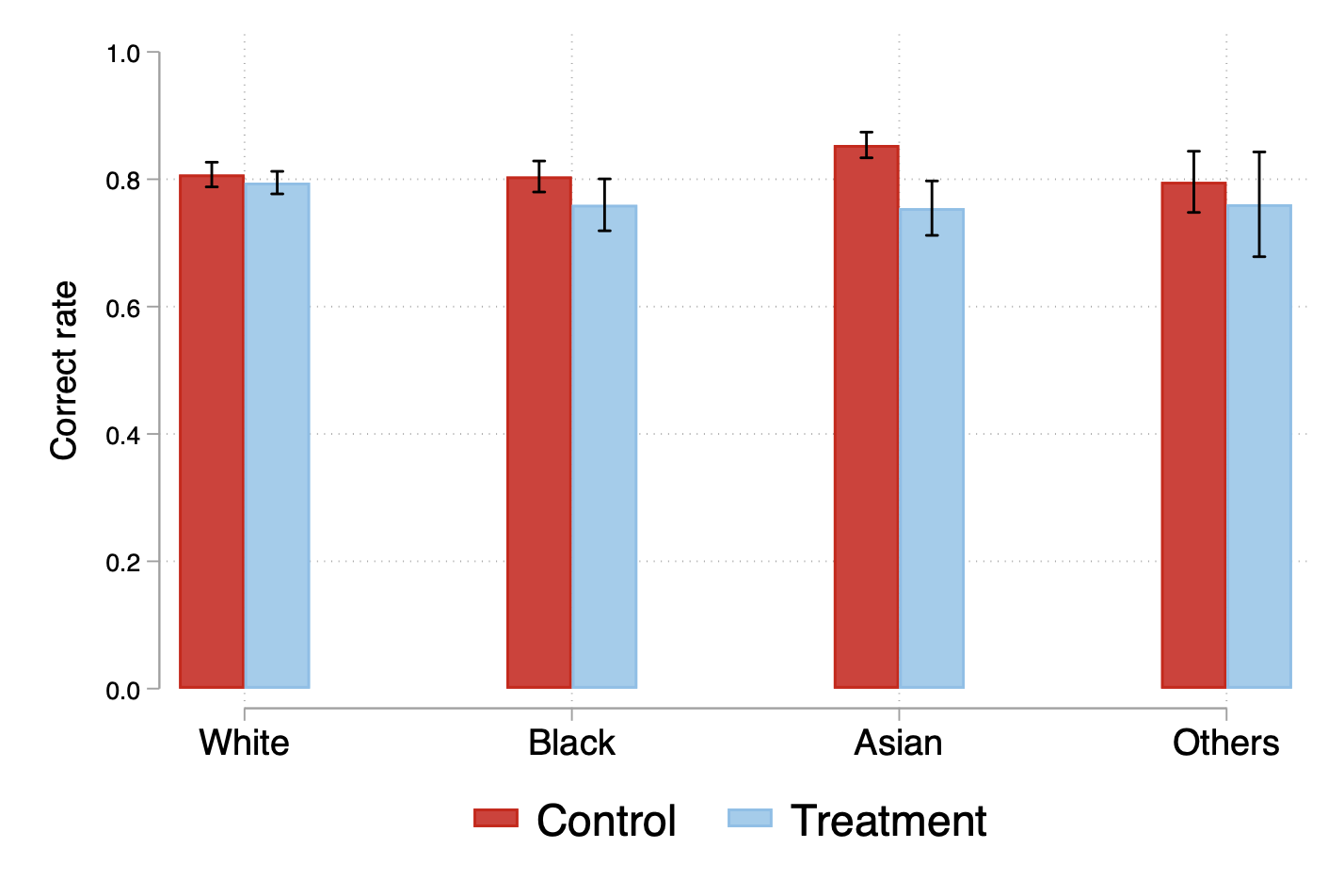}
\caption{Correct answers.}
\end{subfigure}

    \vspace{0.5em}
    \begin{minipage}{0.9\textwidth} 
        \footnotesize \textit{Notes}: Image-specific fixed effects are included. Standard errors are clustered at the participant level.
    \end{minipage}
\label{fig:Results_Ethnicity}
\end{figure}

Figure \ref{fig:Results_Ethnicity} corroborates that the heterogeneous treatment effects are primarily driven by minority participants. Moreover, all non-white ethnic groups appear to contribute to the overall effect. In contrast, white participants in the sample showed no response to the image concerns induced by the treatment. This result provides suggestive evidence that minority participants may be more vulnerable to image-related pressures that discourage AI adoption. For example, in the control group, Asian participants were the most frequent users of AI and achieved the highest accuracy as a group. However, in the treatment group, they became the least likely to use AI, which resulted in the lowest accuracy among all groups. Findings like this carry important policy implications. If certain minority groups, such as Asians in this case, are more susceptible to image concerns, we may be observing a labor market dynamic where success depends not on skill, but on a willingness to ignore social scrutiny. This could disadvantage precisely those groups that are most sensitive to being judged, ultimately reinforcing existing inequities.

\subsection{Mechanisms}
Finally, to shed light on the mechanism, participants were asked two non-incentivized questions, answered on a 5-point Likert scale ranging from ``strongly disagree'' to ``strongly agree'', before joining the video call.

\begin{figure}[h!]
\centering
\caption{\textit{I feel uncomfortable about the upcoming Zoom call.}}
\includegraphics[width=5in]{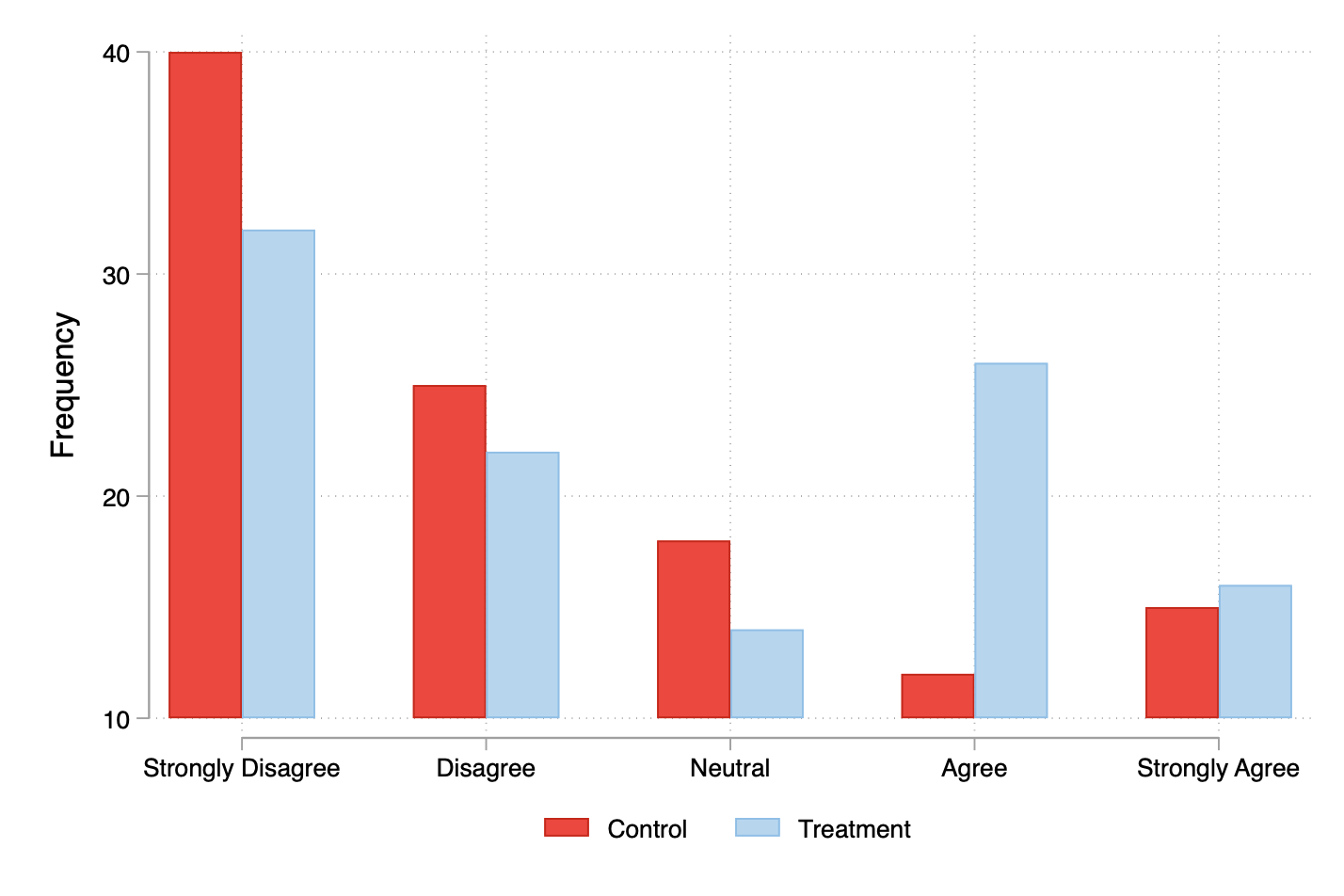}
\label{fig:4}
\end{figure}

First, participants were asked to report how uncomfortable they felt about the upcoming Zoom call. Pooling the two highest levels of discomfort (``Agree'' and ``Strongly Agree''), Figure \ref{fig:4} shows a 55\% increase in participants who felt uncomfortable about the call in the treatment group (42 compared to 27 in the control group). This finding supports psychological discomfort as a likely mechanism behind the treatment effects.

\begin{figure}[h!]
\centering
\caption{\textit{I believe I made good use of AI recommendations during this task.}}
\includegraphics[width=5in]{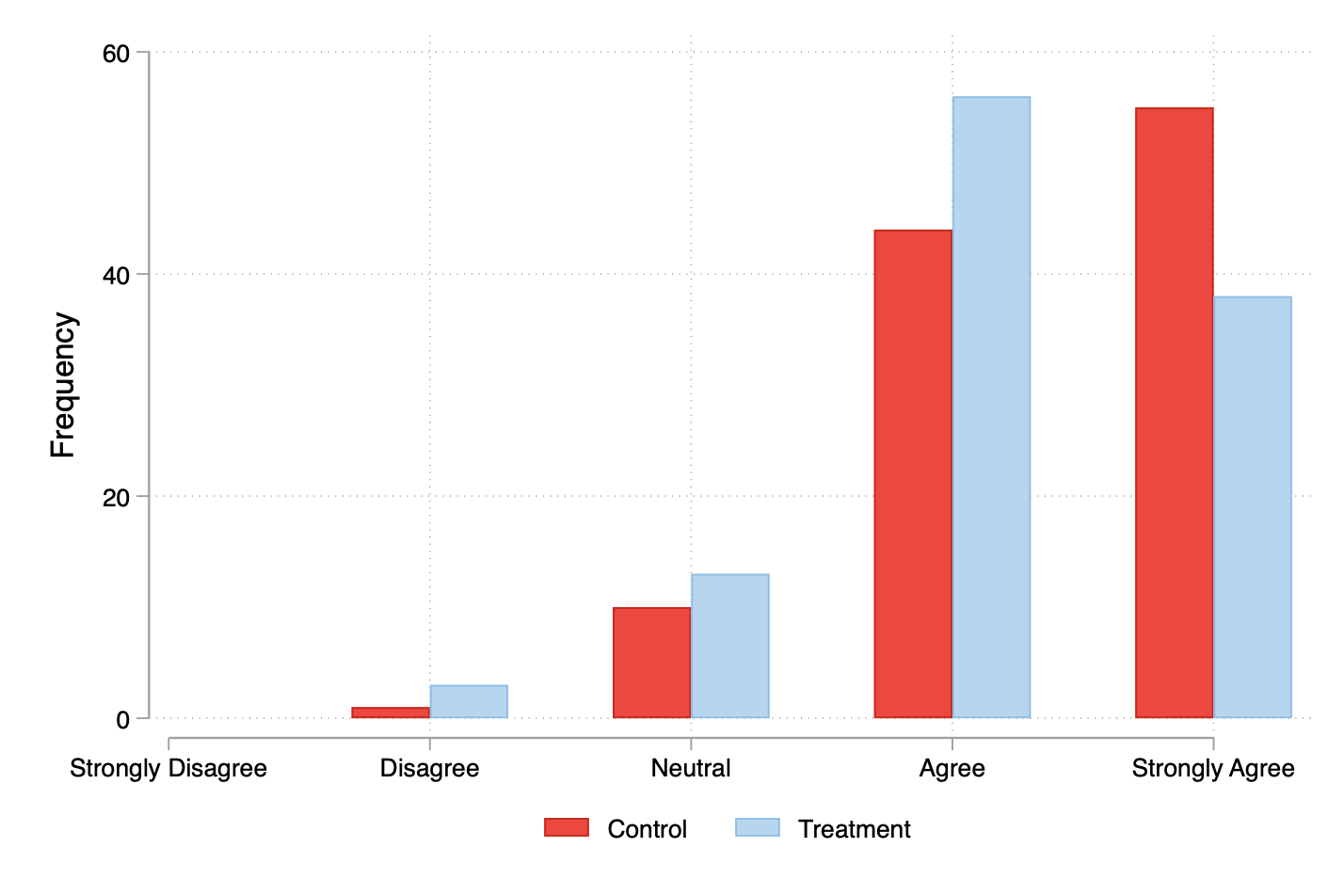}
\label{fig:5}
\end{figure}

The second question explored participants' beliefs about whether they thought they had made good use of AI recommendations during the task. This question was intended to assess whether they felt restricted in choosing to follow AI recommendations in the treatment group, which Figure \ref{fig:5} supports. In the control group, 55 participants strongly agreed that they had made good use of AI, whereas this number decreased by 31\% in the treatment group, with only 38 participants strongly agreeing. This question also hints at a certain level of sophistication among participants, as they were aware of the self-imposed constraints on how they used the AI recommendations.

\section{Conclusion}\label{sec:Con}
As AI technologies become increasingly integrated into professional decision-making, understanding the barriers to effective human-AI collaboration is essential. This paper demonstrates that even in the absence of economic or reputational incentives, subtle image concerns can lead individuals to reject accurate AI recommendations, ultimately reducing performance. The findings show that even modest image concerns, when introduced in a controlled setting, can meaningfully affect behavior. In higher-stakes environments, where social and reputational pressures are expected to be more intense, these effects are likely to be even stronger.

Importantly, the effects of image concerns are not the same for everyone. The treatment had a stronger impact on minority participants, suggesting that even seemingly neutral AI recommendation systems can unintentionally reinforce inequality when social dynamics are overlooked. Much of the \textit{algorithmic fairness} literature has focused on adjusting algorithm outputs \citep{Liang2025}. However, these results highlight the importance of understanding how people interact with algorithms, and how social background can shape those interactions. This extends the call for a broader approach to algorithmic fairness \citep{Ge2025}, one that considers not just the algorithm’s output, but the final outcomes that emerge from human–AI collaboration.

Moving forward, developers, managers, and policymakers should consider designing interventions that explicitly address image concerns. Existing research demonstrates that a clear understanding of these concerns can improve policy outcomes \citep{Butera2022, Jee2024}. To fully realize the potential of human–AI collaboration, it is crucial to consider not only the performance of algorithms but also the social and psychological contexts in which these technologies are implemented.

\clearpage
\bibliographystyle{plainnat} 
\bibliography{bibliography}  
\clearpage

\appendix
\counterwithin{figure}{section}
\counterwithin{table}{section}
\section{Experiment Implementation}

\subsection{Task Interface}\label{TaskInterface}
\begin{figure}[h!]
\centering
\caption*{Initial Choice}
\includegraphics[width=4.5in]{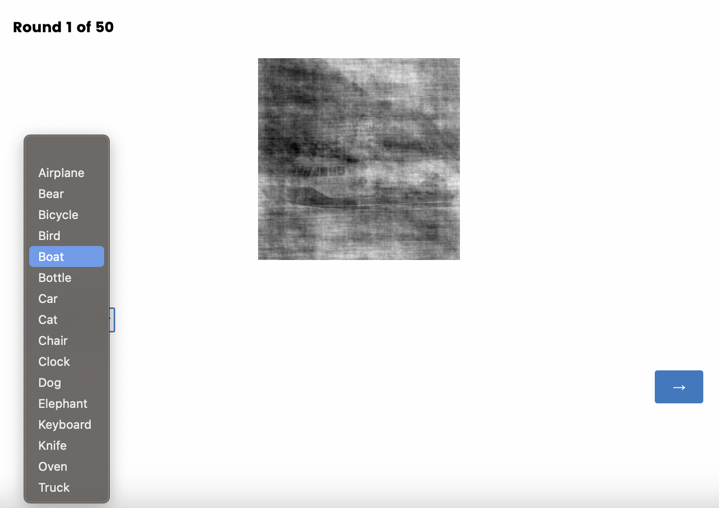}
\end{figure}

\vspace{1cm}

\begin{figure}[htbp]
\centering
\caption*{AI Recommendation Stage}
\begin{subfigure}[b]{.49\textwidth}
\centering
\includegraphics[width=3.2in]{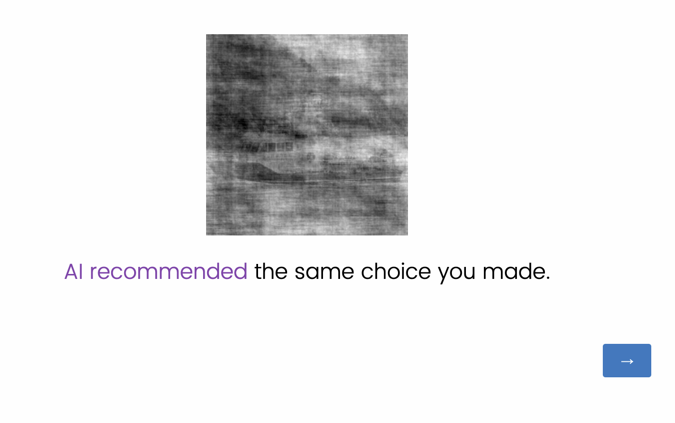}
\caption*{When the recommendation agrees.}
\end{subfigure}
\begin{subfigure}[b]{.49\textwidth}
\centering
\includegraphics[width=3.2in]{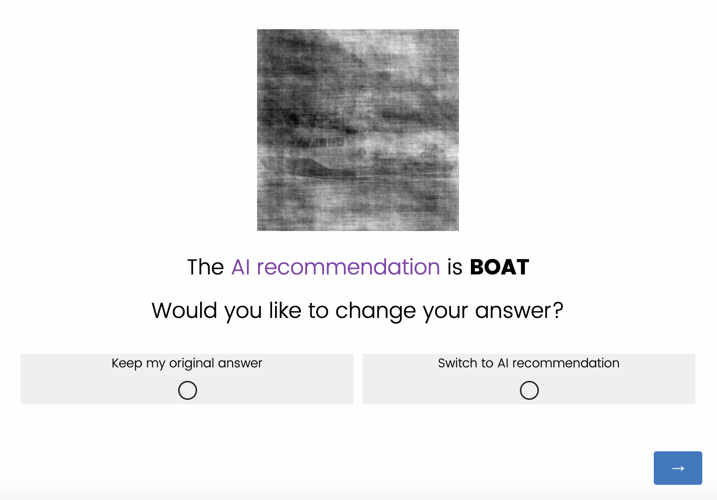}
\caption*{When the recommendation differs.}
\end{subfigure}
\end{figure}
\addtocounter{figure}{-1}

\clearpage
\subsection{Task Design} \label{sec:TaskAppendix}
I retrieved 80 images from the publicly available dataset used by \citet{Steyvers2022}. These images were selected through multiple rounds of piloting to ensure they were diverse in difficulty yet sufficiently challenging.

\begin{figure}[h!]
\centering
\caption{Initial Choice Correct Answers.}\label{fig:TaskDifficulty}
\includegraphics[width=4.2in]{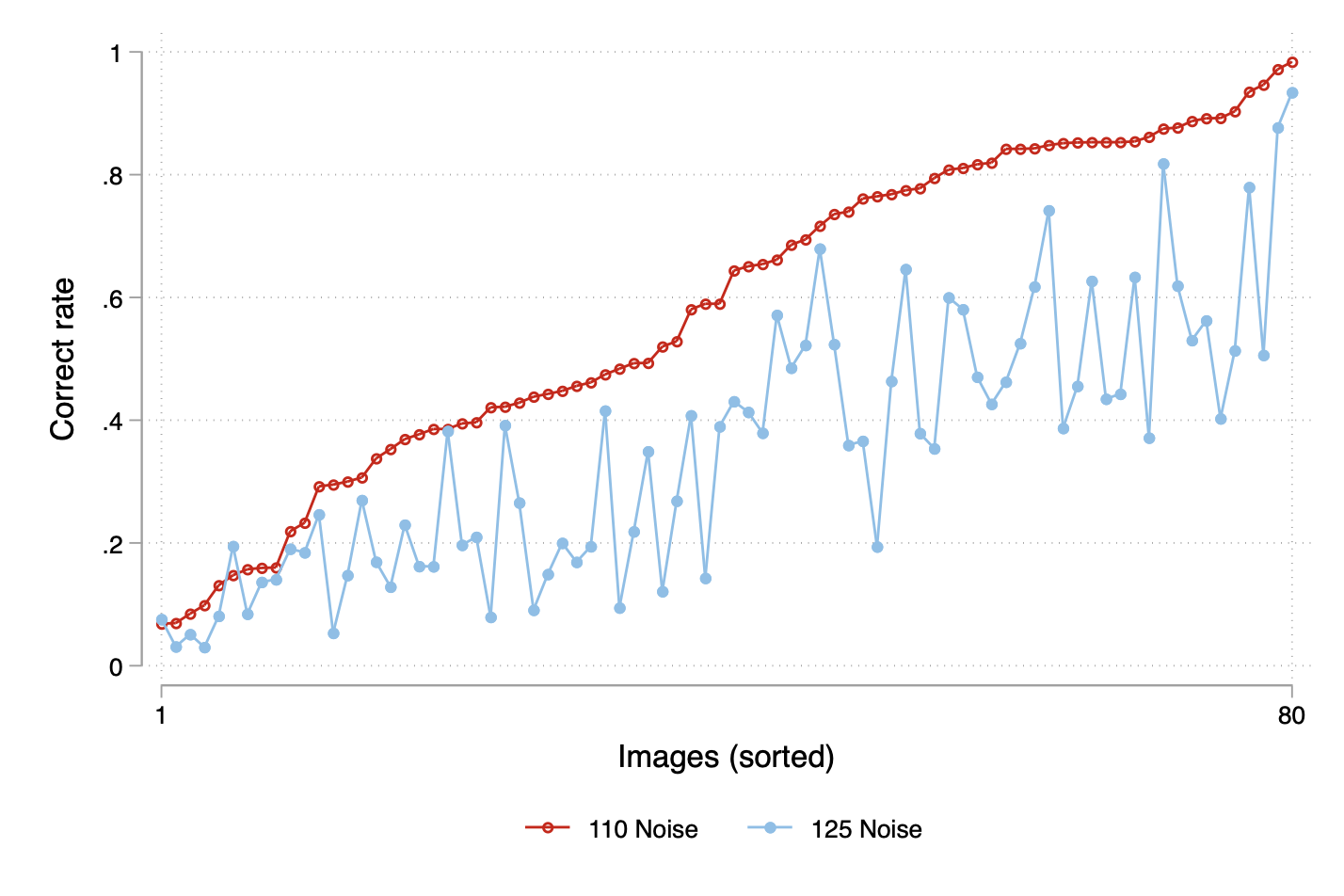}
\vspace{0.5em}
    \begin{minipage}{0.9\textwidth} 
        \footnotesize \textit{Notes}: I compute the accuracy in the first choice as a proxy for image difficulty. Images are then sorted based on this difficulty metric using the 110 noise level (shown in red). The overlaid blue line connects the corresponding accuracy values for each image at the 125 noise level.
    \end{minipage}
\end{figure}

To vary the degree of difficulty in the human and machine classifier experiments, \citet{Steyvers2022} applied four levels of phase noise distortion to the images (80, 95, 110, and 125), which determined the degree of blurriness. This introduced an objective source of variation in image difficulty. In my experiment, I used only the two highest noise levels (110 and 125), with the noise level randomized before an image was shown. Figure \ref{fig:TaskDifficulty} illustrates the variation in task difficulty and highlights a clear within-image increase in difficulty as noise levels rise.

A major advantage of this dataset is that it includes predictions from five pre-trained ImageNet models, along with performance evaluations in collaborative settings: human–machine, two humans, and two machines. This feature enables a non-deceptive experimental design and allows the experiment to draw on algorithms previously tested in studies of human–AI complementarity. In the experiment, I provided participants with predictions from the VGG-19 model, a convolutional neural network with 19 layers. This model was selected because, according to the results in \citet{Steyvers2022}, it exhibited the highest complementarity potential among the available options. However, none of the findings in this paper hinge on the use of this specific model.

\clearpage
\subsection{Experiment Instructions} \label{Instructions}

\begin{figure}[h!]
\centering
\caption*{Recruitment Teaser}
\includegraphics[width=6in]{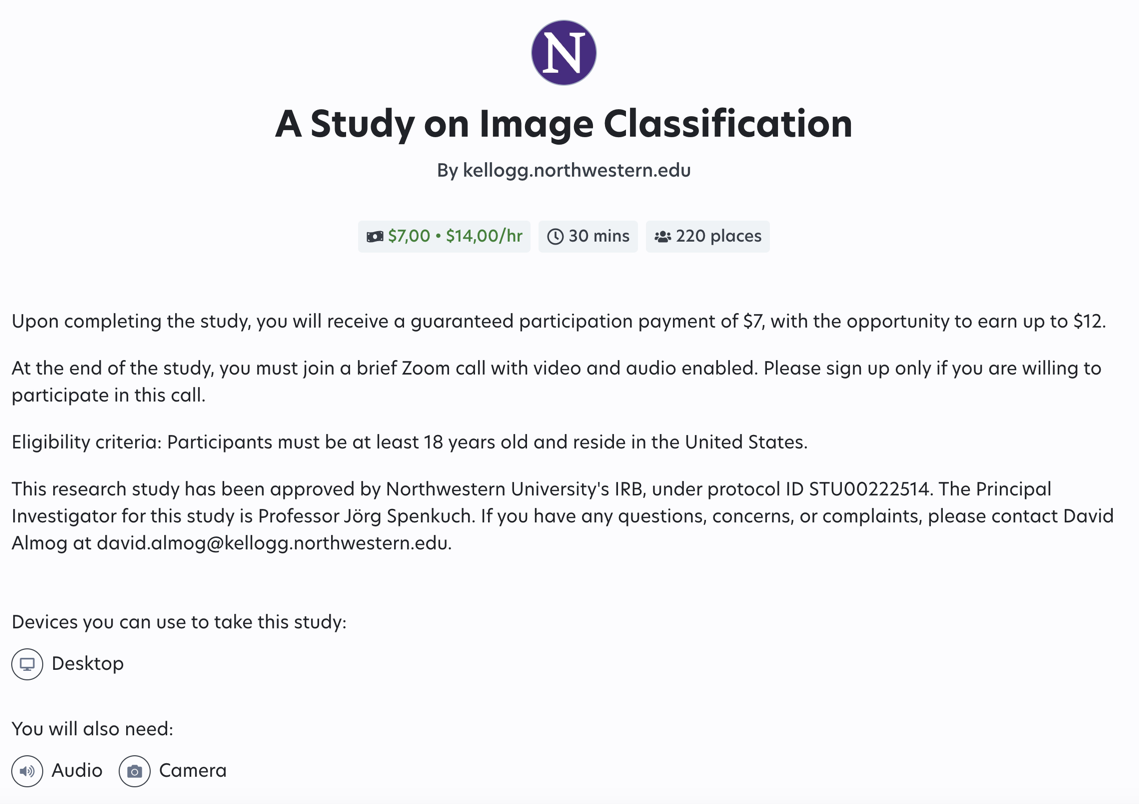}
\label{fig:ProlificRecruitment}
\end{figure}

\begin{figure}[h!]
\centering
\caption*{Instructions Page 1/6}
\includegraphics[width=5.5in]{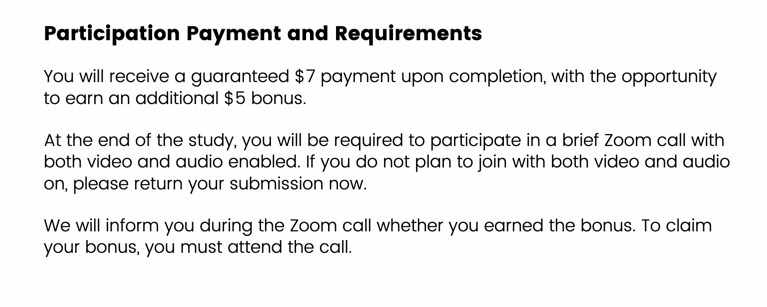}
\end{figure}

\begin{figure}[h!]
\centering
\caption*{Instructions Page 2/6}
\includegraphics[width=5.5in]{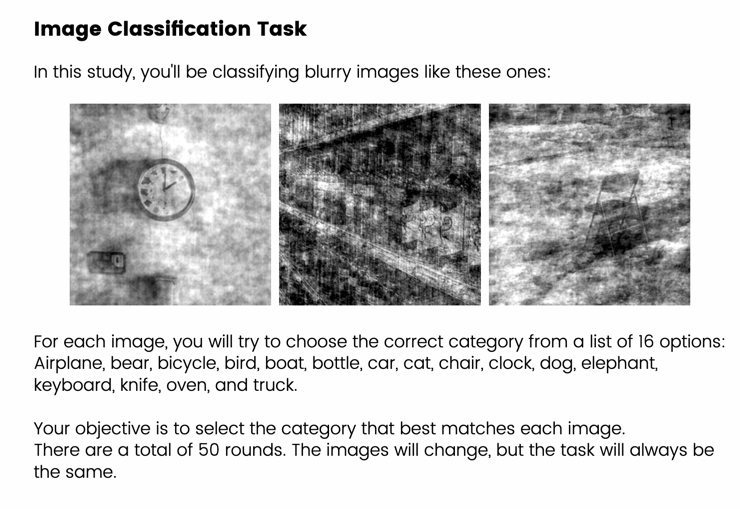}
\end{figure}

\begin{figure}[h!]
\centering
\caption*{Instructions Page 3/6}
\includegraphics[width=5.5in]{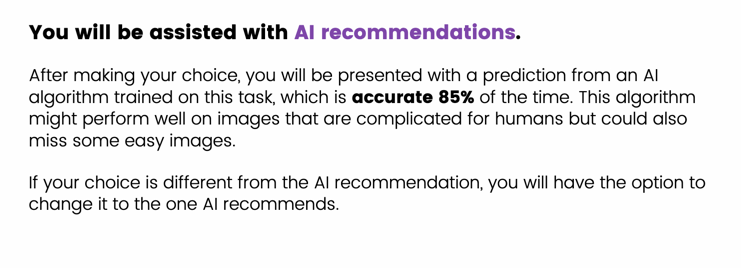}
\end{figure}

\begin{figure}[h!]
\centering
\caption*{Instructions Page 4/6}
\includegraphics[width=5.5in]{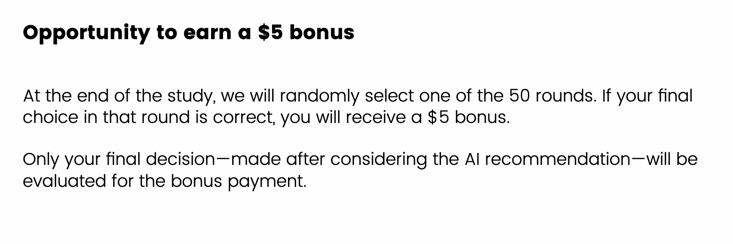}
\end{figure}

\begin{figure}[h!]
    \centering
\caption*{Instructions Page 5/6: Zoom Call Explanation by Experimental Condition}
    \begin{subfigure}[b]{0.85\textwidth}
        \centering
        \includegraphics[width=\textwidth]{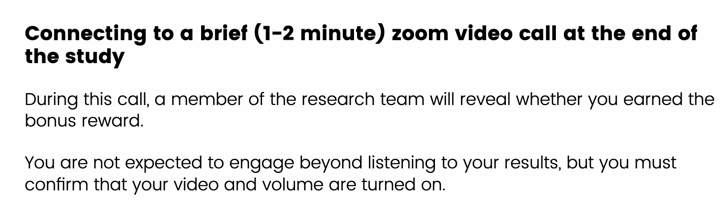}
        \caption{Control group}
    \end{subfigure}
    \begin{subfigure}[b]{0.85\textwidth}
        \centering
        \includegraphics[width=\textwidth]{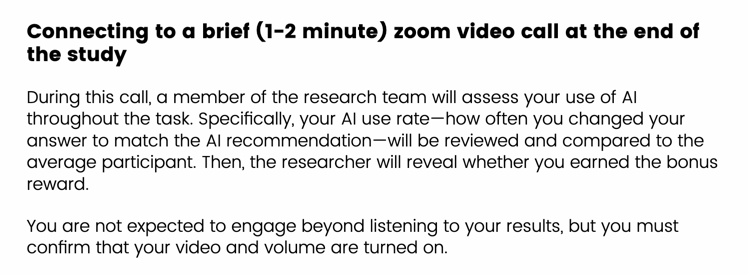}
        \caption{Treatment group}
    \end{subfigure}
\end{figure}

\begin{figure}[h!]
    \centering
\caption*{Instructions Page 6/6: Zoom Call Preview Image by Experimental Condition}
    \begin{subfigure}[b]{0.85\textwidth}
        \centering
        \includegraphics[width=\textwidth]{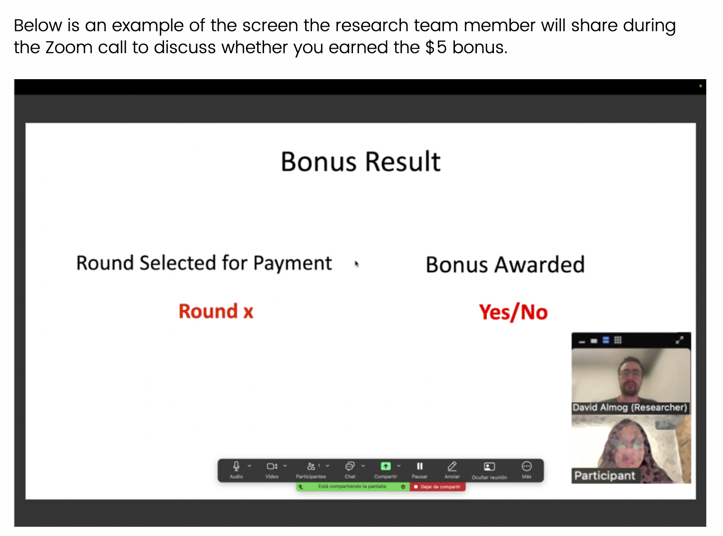}
        \caption{Control group}
    \end{subfigure}
    \begin{subfigure}[b]{0.85\textwidth}
        \centering
        \includegraphics[width=\textwidth]{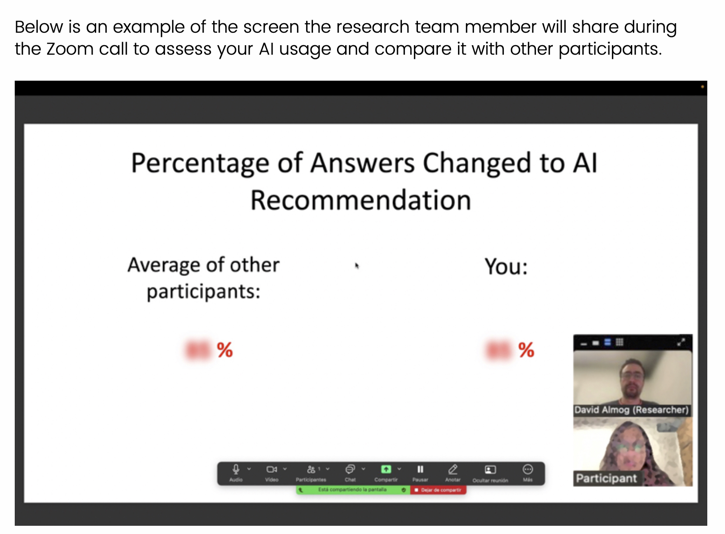}
        \caption{Treatment group}
    \end{subfigure}
\end{figure}

\clearpage
\counterwithin{figure}{section}
\counterwithin{table}{section}
\section{Additional Tables and Figures}
\begin{figure}[htbp]
\centering
\caption{AI Recommendation use by 10-round blocks.}
\begin{subfigure}[b]{.49\textwidth}
\centering
\includegraphics[width=3.2in]{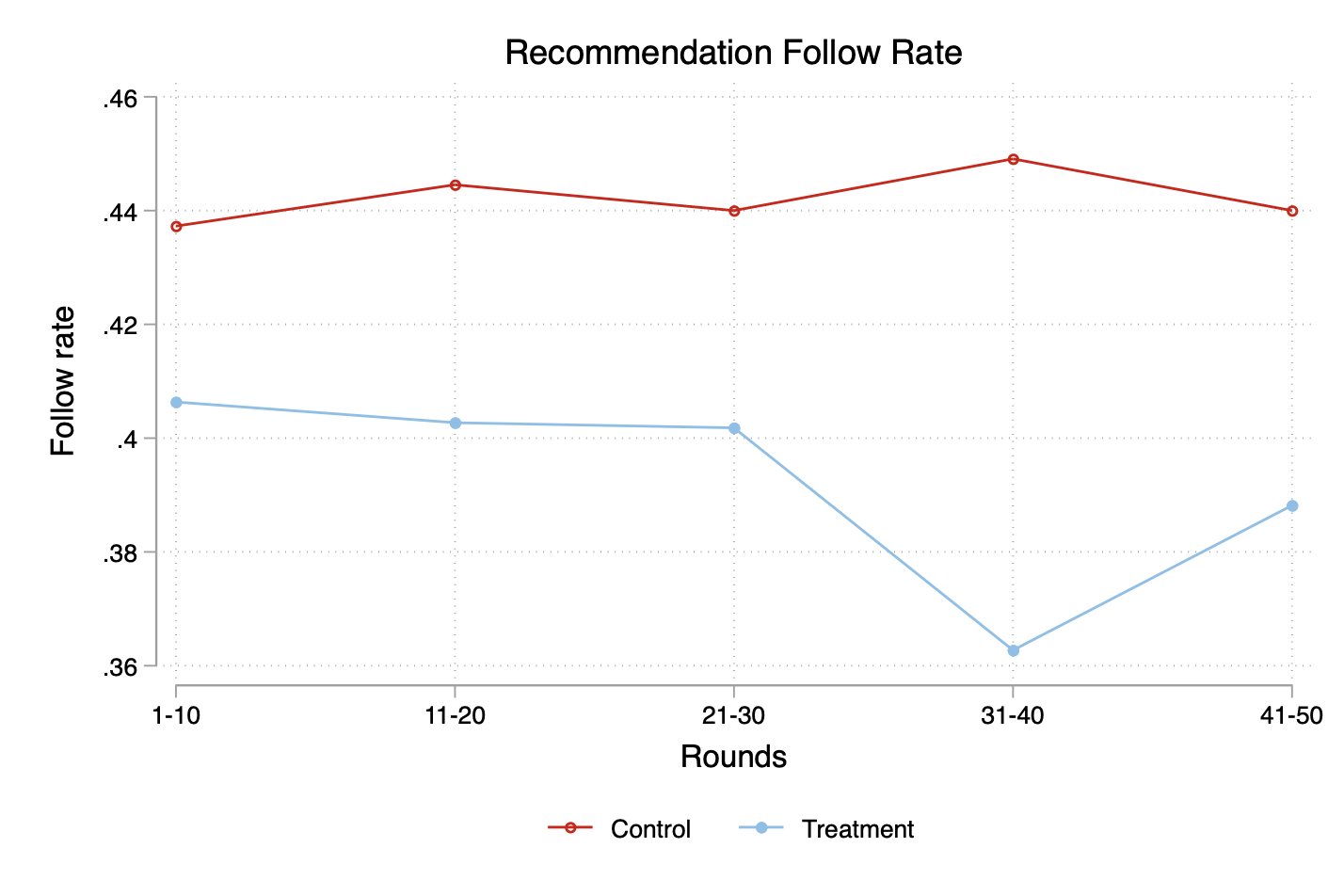}
\caption{All rounds.}
\end{subfigure}
\begin{subfigure}[b]{.49\textwidth}
\centering
\includegraphics[width=3.2in]{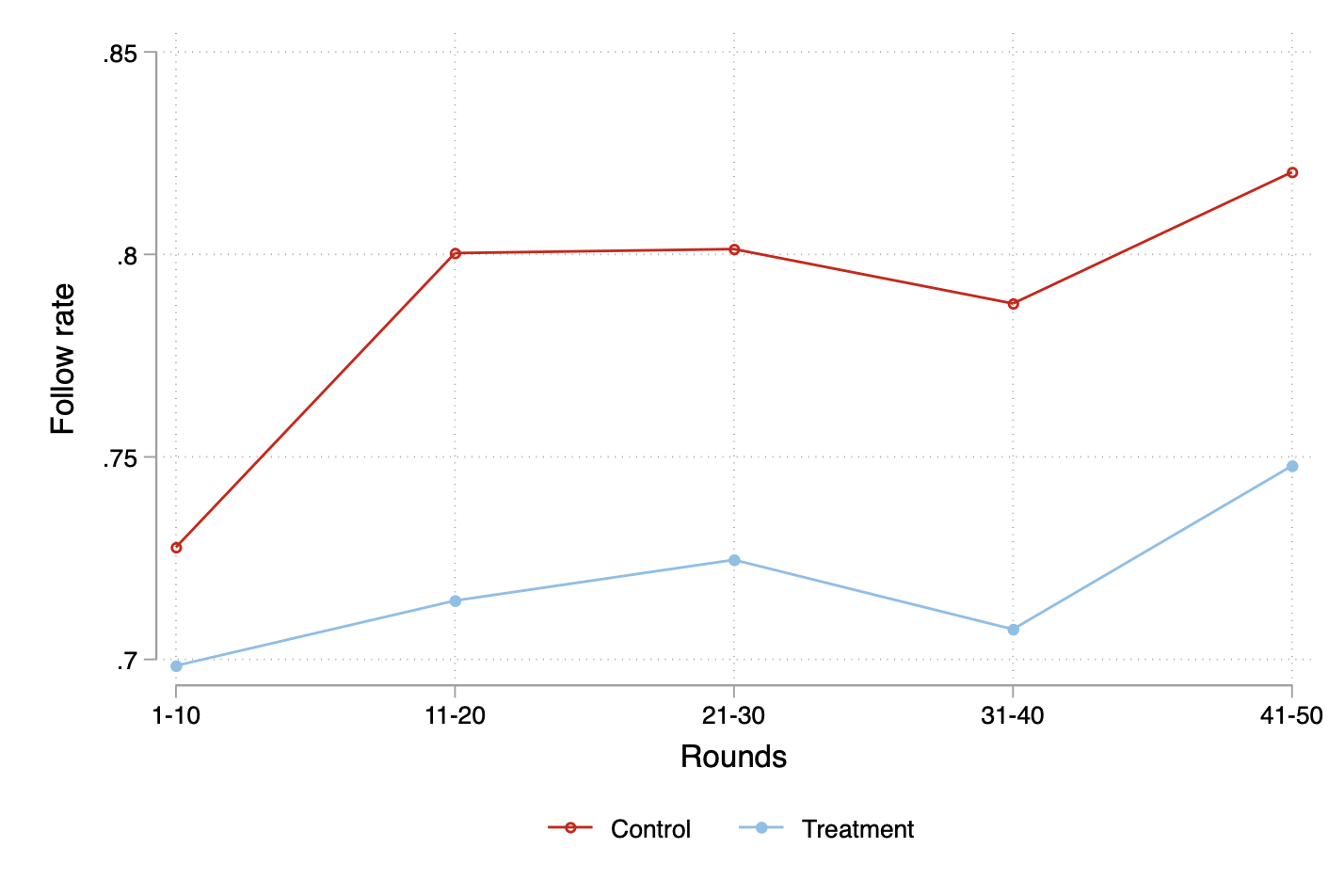}
\caption{Conditional on switching available.}
\end{subfigure}
\label{fig:10rounds}
\end{figure}

\begin{table}[h!]
    \centering
    \caption{Heterogeneous Treatment Effects.}
    \label{tab:Results_Heterogeneity}
    \sisetup{table-format=1.3}  
    \begin{tabular}{lSSSSSSSSS}
        \toprule
        & \multicolumn{2}{c}{AI Recommendation use} & & \multicolumn{1}{c}{Correct answer} & &  \\
        \cmidrule(lr){2-3} \cmidrule(lr){5-5} 
        & {All} & {Conditional} & & {Final}  \\
        & {(1)} & {(2)} & & {(3)}   \\
        \midrule
        {\textbf{Panel A: Gender}} \\
        {Treatment} & {-0.052}* & {-.070}** & & {-0.021} \\
        & {(0.030)} & {(0.035)} & & {(0.015)} \\
        {Male} & {-0.011} & {0.017} & & {0.020}  \\
        & {(0.031)} & {(0.035)} & & {(0.014)} \\
        {\textbf{Treatment*Male}} & \textbf{0.014} & \textbf{0.004} & &\textbf{-0.011} \\\vspace{1em}
        & {(0.043)} & {(0.052)} & & {(0.021)} \\
        Constant & 0.445 & 0.778 & &  0.799 \\
        \midrule
        {\textbf{Panel B: Age}} \\
        {Treatment} & {-0.016} & {-0.052} & & {-0.025}* \\
        & {(0.029)} & {(0.037)} & & {(0.015)} \\
        {Older} & {0.080}*** & {0.053} & & {0.004}  \\
        & {(0.030)} & {(0.035)} & & {(0.014)} \\
        {\textbf{Treatment*Older}} & \textbf{-0.056} & \textbf{-0.029} & &\textbf{-0.004} \\\vspace{1em}
        & {(0.042)} & {(0.051)} & & {(0.022)} \\
        Constant & 0.398 & 0.757 & & 0.807  \\
        \midrule
        {\textbf{Panel C: Ethnicity}} \\
        {Treatment} & {-0.008} & {-0.026} & & {-0.013} \\
        & {(0.027)} & {(0.034)} & & {(0.013)} \\
        {Minority} & {0.062}** & {0.036} & & {0.004}  \\
        & {(0.030)} & {(0.034)} & & {(0.014)} \\
        {\textbf{Treatment*Minority}} & \textbf{-0.096}** & \textbf{-0.111}** & &\textbf{-0.040}* \\\vspace{1em}
        & {(0.044)} & {(0.052)} & & {(0.023)} \\
        Constant & 0.415 & 0.771 & & 0.807  \\
        \midrule
        Observations & {11,000} &  {6,098} &  & {11,000}  \\
        \bottomrule
    \end{tabular}
    
    \vspace{0.5em}
    \begin{minipage}{0.9\textwidth} 
        \footnotesize \textit{Notes}: Each dependent variable is regressed on a treatment indicator (equal to 1 if the participant was assigned to the treatment group), a group indicator (equal to 1 if the participant is male, older, or a minority—depending on the panel), and the interaction between the two. Image-specific fixed effects are included. Standard errors are clustered at the participant level.\\
        * \(p<0.10\), ** \(p<0.05\), *** \(p<0.01\).
    \end{minipage}
\end{table}

\end{document}